\documentclass[preprint,12pt]{elsarticle}

\usepackage{amssymb}

\begin{document}

\begin{frontmatter}



\title{Pathlength scaling in graphs with incomplete navigational information}


\author[umu]{Sang Hoon Lee\corref{cor}}
\ead{sanghoon.lee@physics.umu.se}

\author[umu,skku,su]{Petter Holme}

\cortext[cor]{Corresponding author}

\address[umu]{IceLab, Department of Physics, Ume{\aa} University, 901 87 Ume{\aa}, Sweden}
\address[skku]{Department of Energy Science, Sungkyunkwan University, Suwon 440--746, Korea}
\address[su]{Department of Sociology, Stockholm University, 106 91 Stockholm, Sweden}

\begin{abstract}
The graph-navigability problem concerns how one can find as short paths as possible between a pair of vertices, given an incomplete picture of a graph. We study the navigability of graphs where the vertices are tagged by a number (between 1 and the total number of vertices) in a way to aid navigation. This information is too little to  ensure errorfree navigation but enough, as we will show, for the agents to do significantly better than a random walk. In our setup, given a graph, we first assign information to the vertices that agents can utilize for their navigation. To evaluate the navigation, we calculate the average distance traveled over random pairs of source and target and different graph realizations. We show that this type of embedding can be made quite efficiently; the more information is embedded, the more efficient it gets. We also investigate the embedded navigational information in a standard graph layout algorithm and find that although this information does not make algorithms as efficient as the above-mentioned schemes, it is significantly helpful.
\end{abstract}

\begin{keyword}
Graph routing \sep
Geometric routing \sep
Navigability


\end{keyword}

\end{frontmatter}


\section{Introduction}
\label{introduction}
Navigability or searchability---how agents can find their way from a source to a target on a graph---has been a vivid subfield of network studied in the last decade. Different studies have focused on different situations regarding the amount of information accessible to the agents. Some studies investigate degree-biased random walks~\cite{Adamic2001}; others assume that agents have geographic information~\cite{Kleinberg2000,Boguna2008,Boguna2008a}. Yet other works assume that the agents obtain the direction along the way and focus on the information content needed, per vertex, to find the shortest path~\cite{MartinRosvall}. The scenario in this work, similar to Ref.~\cite{Holme2007a}, is that a limited amount of information can be distributed to the vertices and exploited by the agents to aid their navigation.
The problem of finding efficient paths for a packet or a piece of information
from a certain source to a target is not only  mathematically
intriguing, but also directly related to practical
applications  where the knowledge about the entire system is usually not accessible. Examples of such systems include Internet routing protocols~\cite{TraceRoute1,TraceRoute2}
or road networks~\cite{HYoun2007}.

In this paper we investigate efficient ways, given the graph structure, to distribute limited information (in the form of unique indices) to the vertices and assign a protocol to the agents to interpret this information, for the benefit of navigation. We assume the agents have a memory of where they have been, but no \textit{a priori} knowledge of the graph.
This scenario was first discussed in
Ref.~\cite{Holme2007a}, where the problem was defined and some simple strategies were proposed. Here we improve the protocols of Ref.~\cite{Holme2007a} and address questions about the amount of added information and the performance of the navigation.
Furthermore, we investigate how much the geographic information (assigned from a spring-embedding algorithm), helps the mentioned navigation-by-vertex-indices problem.

We emphasize that the navigation processes in this work focus on the {\em user} of networks, rather than the designer or the architect of the system. In other words, we are interested in the efficiency of the navigators' strategies based on information, not the efficiency of information embedding in the construction stage. This subject, we believe, is a very {\em practical} approach in this respect, because the situation when facilities are actually used is our main interest. For instance, a mobile ad hoc network (MANET) is a self-configuring network of mobile devices which continuously maintains the information~\cite{MANET}. Another example is a wireless sensor network (WSN) consisting of spatially distributed autonomous sensors to cooperatively pass their data through the network to a main location~\cite{WSN}. These examples illustrate how important the problem of allocating information to users or devices is in the digital era. The rest of the article presents a more detailed definition of the index-based navigation protocol with the simulation results, and the incorporation of the geometry-based navigation. A discussion of the relation of information and efficiency of navigation algorithms concludes our work.

\section{Preliminaries}
\label{Preliminaries}

Our starting point is a graph $G=(V,E)$ where $V$ is the set of $N$ vertices and $E$ is the set of $M$ edges.
The basic idea is that one assigns information 
to each vertex. The information associated to the agent's current vertex and that to its neighbors is accessible to the agent. An agent is also equipped with a strategy to choose the next vertex to visit based on cumulated information about the vertices already visited in her memory, the indices of the current neighbors, and the knowledge of the method to assign indices to the vertices.
Most of the simulations will evaluate the efficiency of the indexing scheme and the navigation by measuring the average number of edges covered on the way from source to target.
We consider only connected graphs to make sure
that there is always a path between any pair of two vertices.

\section{Routing by embedded indices}
\subsection{The ASU and ASD protocols}

\begin{figure}
\begin{center}
\includegraphics[width=0.7\textwidth]{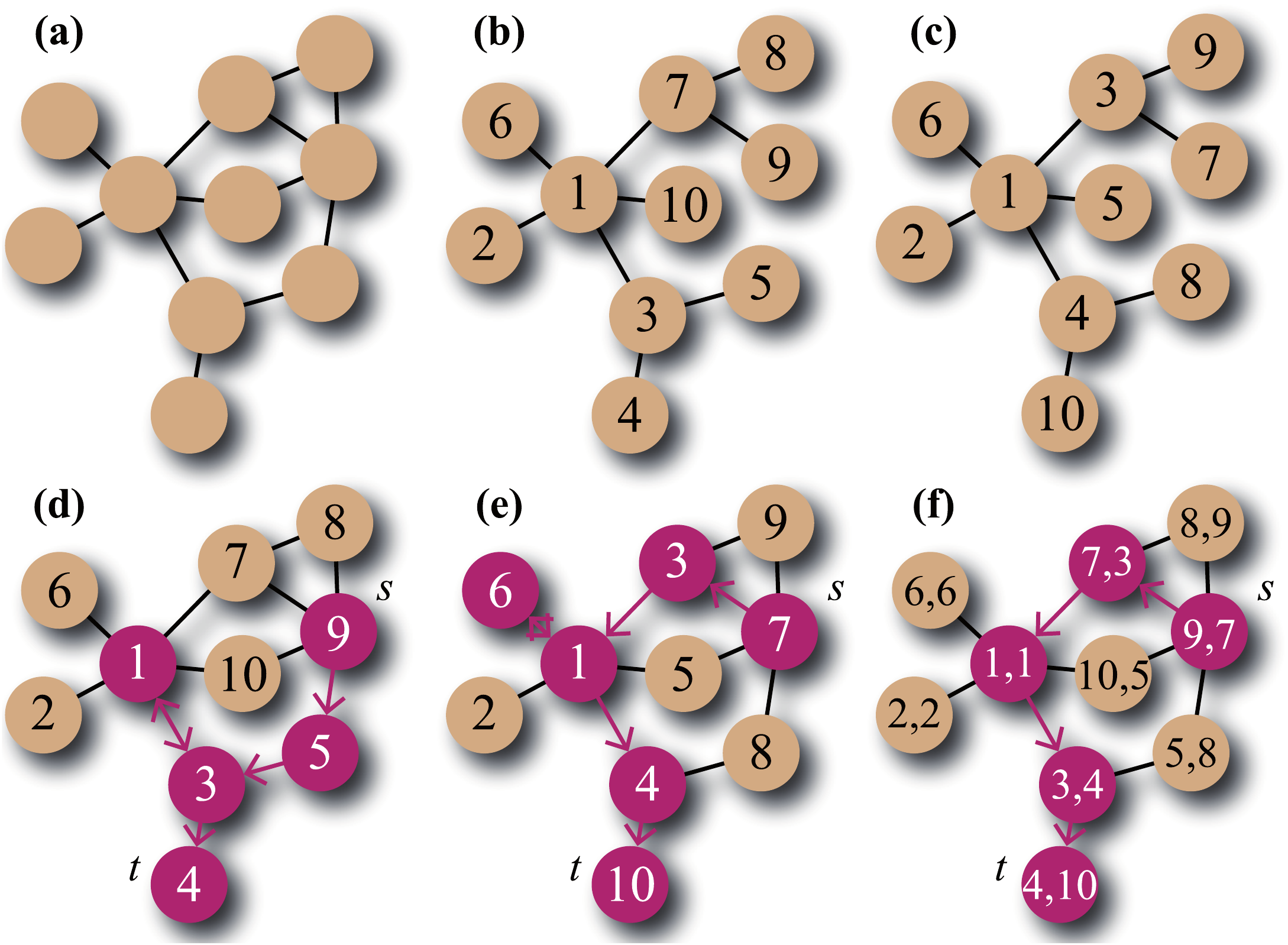}
\caption{Illustration of routing examples
of an indexed graph. (a) The original graph $G$. (b) The spanning
tree $T$ from the root vertex `1'
with ASD indices.
(c) The same $T$ as (b) with ASU indices.
(d) An example routing with ASD indices from the source $s=9$
to the target $t=4$.
(e) An example routing with ASU indices from $s=7$ to $t=10$.
(f) An example routing with ASUD indices from $s=9,7$ to $t=4,10$,
where the numbers next to the vertices represent ASD and ASU indices respectively''.
}
\label{IndexedGraphExample}
\end{center}
\end{figure}

The first step in the definition of the two protocols---\textit{accurate search up} (ASU) and \textit{accurate search down} (ASD)---is to identify one vertex as \textit{root} with the first index `$1$.' On a path $s \to t$, the root will always be passed, so by the terminology from search trees, all paths will go up toward the root, then down toward the target.
For the methods to be efficient, 
$1$ is chosen as a vertex of minimal eccentricity (maximal distance to any other vertex). After $1$ is found, one constructs a spanning tree $T$ such that the distances from $1$ is the same as in $G$, as shown in Fig.~\ref{IndexedGraphExample}(c). In this example, the tree $T$ is composed of the root $1$, the vertices with the distance one from $1$: $\{ 2, 3, 4, 5, 6 \}$, and the vertices with the distance two from $1$: $\{ 7, 8, 9, 10 \}$ [Fig.~\ref{IndexedGraphExample}(c)]. For ASD [Fig.~\ref{IndexedGraphExample}(b)] we assign the indices to the vertices of $G$ from $1$ to $N$ in their order of appearance in a depth-first search (DFS) from $1$ on $T$, i.e., the indices are consecutively assigned for each ``branch'' of $1$; for instance, vertices $\{3, 4, 5\}$ correspond to the branch set from $3$. The indexing of the ASU scheme [Fig.~\ref{IndexedGraphExample}(b)] is basically the order of appearance in a breadth-first search (BFS) from $1$, which is already assigned in constructing $T$ as we have explained above (details in \cite{Holme2007a}).
Here, once again, we emphasize that even though the calculation of eccentricities is not an efficient process, the eccentricity calculation, determination of $1$, and the construction of $T$ is the process only necessary once at the very beginning of the routing process. In other words, the information embedded in the indices are usable as long as the structure of the graph is maintained.

The navigation strategy of ASD and ASU is first moving toward indices as low as possible until $1$ is reached, then going toward the vertex with the largest index smaller than $t$ until $t$ is reached. One can show that for ASD (ASU) indices the way $1\to t$ ($s \to 1$) is guaranteed to be the shortest, but $s\to 1$ ($1 \to t$) could be longer, respectively~\cite{Holme2007a}. In the ASU scheme, the agents navigate by first moving toward lower indices, then (when $1$ is reached) moving toward the vertex with an index closest to $t$.
``Backtracking'' based on the navigator's memory is sometimes necessary in this ASU($1 \to t$) route.
Figure~\ref{IndexedGraphExample} illustrates three different
navigation methods for source-target pairs on a simple graph.
In this example, the root vertex $1$ [denoted in Figs.~\ref{IndexedGraphExample}(b) and (c)] is chosen by its smallest value
of eccentricity $=2$. The spanning tree, shown in Figs.~\ref{IndexedGraphExample}(b) and (c),
is indexed with ASD and ASU schemes, respectively. An example of ASD routing is shown in Fig.~\ref{IndexedGraphExample}(d),
where the navigator starting from $s=9$ chooses $5$, the smallest index among her neighbors, and follows the rule
previously described until she reaches $1$. 
Note that there is two-step inefficiency in the $9 \to 1$ route,
compared to the global shortest route $9 \to 10 \to 1$. In an ASU routing shown in Fig.~\ref{IndexedGraphExample}(e),
the route $s(=9) \to 1$ is the real shortest path, while the inefficiency occurs during the $1 \to t(=10)$ route, due to
the ``backtracking'' process $6 \to 1 \to 4$(the neighbor of $1$ whose ASU index is the next closest to $10$).
In general, especially for graphs with broad degree distributions, ASD turns out to be more efficient than ASU~\cite{Holme2007a}.
In ASD, it is likely that an agent reaches a hub, which is a vertex with a large number of connections, quite efficiently even though ASD does not guarantee the $s \to 1$ part, for the reason that a hub has many neighbors~\cite{JDNoh2004}.
In other words, taking the vertex with the minimal index among all the hubs' large number (by definition) of neighbors reduces the search space considerably~\cite{JDNoh2004}.
For ASU the hubs do not help in reducing the search space from $1$ to $t$ in the same way.
One can, however, construct special cases where ASU is more efficient than ASD~\cite{Holme2007a}.

\subsection{ASU+ASD=ASUD}
\label{ASUD}


Increasing the information per vertex indeed lowers the pathlengths. If one allows to put information corresponding to a map of the entire graph (or probably much less than that), the navigation will always be along the shortest paths. The first step from the previous section is to double the information and put both the ASU and ASD index on each vertex. Then agents use ASU to find the root and ASD to the target so the average pathlength will be equal to twice the average shortest distance from the root to other vertices of the graph.
The only inefficiency of this strategy, which we call as ASUD,
comes from the possibility that the actual shortest paths $s \to t$ do not include
$1$. The ASUD scheme in Fig.~\ref{IndexedGraphExample}(f) shows the same
source-target pair as in Fig.~\ref{IndexedGraphExample}(d), but the inefficiency for $s=9,7 \to 1,1$ is removed from
the usage of the ASU indices by choosing $7,3$ in the first step. Finally, note that there still remains one step inefficiency,
because the global shortest path $9,7 \to 5,8 \to 3,4 \to 4,10$ does not include $1,1$ in the path.

If the number of vertices $N_{d}$ for each depth $d$ of the tree from $1$
is known, the average ASU($s \to 1$) pathlength [simply denoted as ASU($s \to 1$) from now on]
is trivially calculated as $\sum_{d=1} d N_{d} / (N-1)$. However, calculating the number of vertices
for each depth is nontrivial and only some limiting cases of
the configuration model~\cite{Newman2001} with the power-law degree distribution with the lower cutoff
degree $k_c = 1$~\cite{Kalisky2006}, are known. This limiting case is not practical for calculating
general cases, so here we briefly argue the relationship among pathlengths and that there exists an {\em upper bound} for ASU($s \to 1$),
which actually corresponds to the pathlength in the thermodynamic limit $N \to \infty$ for homogeneous graphs.
Note that
\begin{equation}
\begin{array}{l}
\textrm{ASUD} = 2\times\textrm{ASD}(1 \to t) = 2\times\textrm{ASU}(s \to 1),
\end{array}
\label{pathlength_relation}
\end{equation}
obviously from the symmetry between ASU and ASD, as verified numerically. An important
conclusion here is that since ASU($s \to 1$) is the average distance between
$1$ and the other vertices with the real shortest path, the upper bound
is given by
\begin{equation}
\textrm{ASUD} \le 2 \bar{l} ,
\label{upper_bound}
\end{equation}
where $\bar{l}$ is the average real shortest path for all the source-target pairs and the
inequality comes from the choice of $1$ with the minimum eccentricity.
In other words, the time complexity of ASUD cannot exceed that of
the real shortest path from global information except for the constant $2$.
In the case of very homogeneous graphs where all the vertices are almost
equivalent, we can expect
\begin{equation}
\lim_{N \to \infty} (\textrm{ASUD})  = 2\bar{l},
\label{ER_graph_limit}
\end{equation}
which will be numerically confirmed later.

\subsection{Simulation results}
\label{IndexedRouting_simulation}

\begin{figure}
\begin{center}
\includegraphics[width=\textwidth]{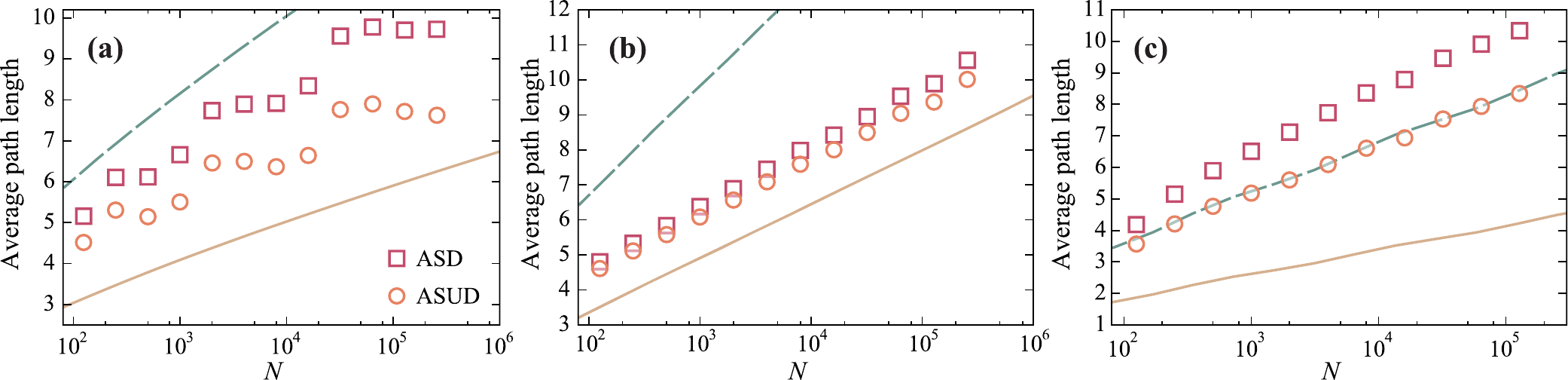}
\caption{Average navigation pathlengths for
routing with indexing, in the case of eccentricity-based root `1': ASD vs ASUD.
The lower and upper lines  represent the real shortest path $\bar{l}$ and $2\bar{l}$ (the upper
bound of ASUD), respectively.
(a) BA model~\cite{Barabasi1999} with $m = 2$,
(b) HK model~\cite{Holme2002b}
with $m = 2$ and a maximal rate of triangles forming, and (c)
ER random graph~\cite{Erdos1959} with the number of edges
$M = 10N$, where $\lim_{N \to \infty} (\textrm{ASUD})  = 2\bar{l}$ holds.
The average pathlengths for ASUD,
$2\times$ASD($1 \to t$), and $ 2\times$ASU($s \to 1$) are separately measured but
the values are essentially the same within the error range,
so only the ASUD case is shown.
Standard error bars would be smaller than the symbols and are not shown.
}
\label{ASD_ASU_real}
\end{center}
\end{figure}

Now we proceed to study the ASUD routing more systematically on standard model graphs.
In Fig.~\ref{ASD_ASU_real} we show results for size scaling of the average $s\to t$ pathlength for the Barab{\'a}si--Albert (BA) model~\cite{Barabasi1999}, the
Holme--Kim (HK) model~\cite{Holme2002b},
and the Erd\H{o}s--R{\'e}nyi (ER)~\cite{Erdos1959}. The ER model simply adds edges between random pairs of vertices until $G$ has $M$ edges. The BA model illustrates the emergent scale--free degree distribution in growing graphs with preferential attachment (the probability that a new vertex connects to an old vertex of degree $k$ is proportional to $k$). The graph is grown by adding one vertex and $m$ edges attached to this vertex, at a time step. The HK model modifies the BA model so that, briefly speaking, the graph contains a nonvanishing ratio of triangles to the maximal possible number of triangles given the set of degrees of the graph.

In our simulations, we
use averages over up to $10^3$ graph realizations and up to $10^4$ source-target pairs per graph. The results, seen in Fig.~\ref{ASD_ASU_real}, show that
ASUD clearly outperforms ASD. We also verify $\textrm{ASUD} = 2\times\textrm{ASD}(1 \to t) = 2\times\textrm{ASU}(s \to 1)$ from the observation that $2\times$ASD($1 \to t$) and $2\times$ASU($s \to 1$) are indistinguishable from ASUD results.
Basically, since all the model graphs considered here show at most the logarithmic dependence of $\bar{l}(N)$ as $N$ increases~\cite{Cohen2003}, the ASUD upper bound shown in Fig.~\ref{ASD_ASU_real} also shows at most logarithmic size dependence.
For the BA graph, ASUD gives about one or two step longer paths than the shortest path over six orders of magnitude with the characteristic ``step'' structure stemming from the discrete increment of minimum eccentricity of $1$. This advantage is smaller for the HK and ER graphs. For the HK model, the combined model gives only a marginal improvement, which implies that ASD finds the root almost as efficient as possible in these graphs. Also, the step structure is absent for the HK model, due to the steady increase of average minimum eccentricity of $1$, where the added edges for clustering enhance the randomness of the graph. For the ER model, the homogeneity of vertices clearly verifies $\textrm{ASUD}  = 2\bar{l}$. For other model parameters, the same qualitative conclusions hold.
Also the combined algorithm improves fairly little on ASU, considering it uses twice the information of the ASD scheme, and the room for improvement is large.

\begin{figure}
\begin{center}
\includegraphics[width=\textwidth]{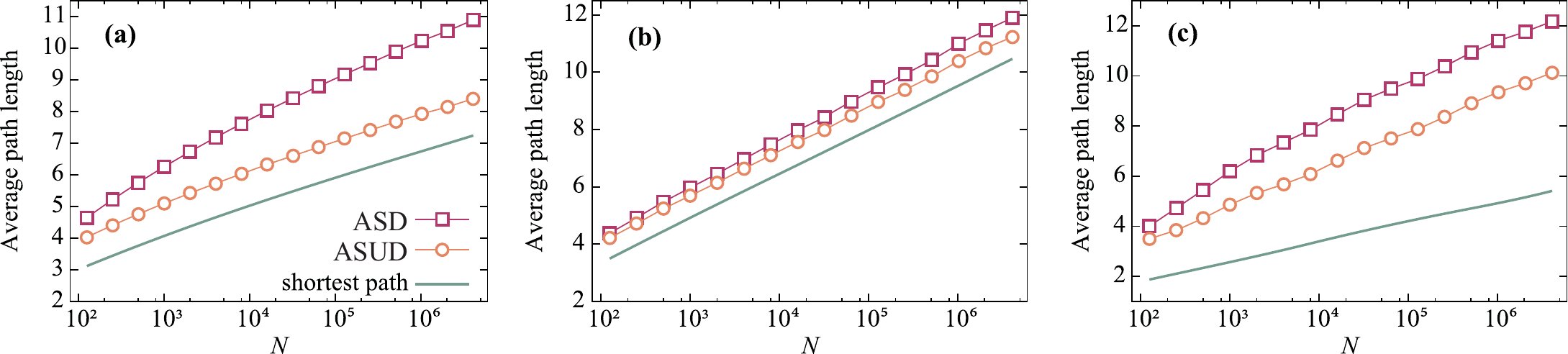}
\caption{Average navigation pathlengths for
routing with indexing, in the case of degree-based root `1': ASD vs ASUD,
for the same graphs and parameter values as in Fig.~\ref{ASD_ASU_real}.
Standard error bars would be smaller than the symbols and are not shown,
and lines are guides to the eyes.
}
\label{ASD_ASU_hub}
\end{center}
\end{figure}

In practice, calculating eccentricity itself can be time-consuming for large graphs (even once). Inspired by the fact that in most cases eccentricities and degrees are highly correlated, we can also choose the vertex of the largest degree as the root and the simulation results are shown in Fig.~\ref{ASD_ASU_hub}. The usage of `degree-based root' is computationally much cheaper, so we can simulate up to $N = 4 \times 10^6$ for the degree-based root case. Furthermore, it often makes the algorithms more efficient than the eccentricity-based root case, surprisingly. The reason that the eccentricity does not perform so well, even if it gives the lowest possible search trees $T$, is that the most eccentric vertices are often a rather large set of vertices not all of which lie on many shortest paths. These results, therefore, illustrate the necessity of distinction between ``bare'' and ``useful'' information from a system, because it is clear that the time complexity of calculating quantities (eccentricity vs. degree) alone cannot necessarily predict the amount of useful information. In this particular case, even if the vertices with the minimum eccentricity guarantee the mathematical upper bound in Eq.~(\ref{upper_bound}), the homogeneity of eccentricity values across the vertices of model graphs do not play a crucial role to enhance the performance. Such homogeneity causes relatively large structural entropy for graph ensembles~\cite{Bianconi2009}, which can be interpreted as lack of useful information. In contrast, the heterogeneity in degrees in scale-free network models are well known and especially degrees are largely correlated with betweenness centralities~\cite{KIGoh2003} representing the centrality in terms of shortest path routing, which yields the useful information available.

\section{Routing by geometric information}
\label{Geometric_information}

\begin{figure}
\begin{center}
\includegraphics[width=0.7\textwidth]{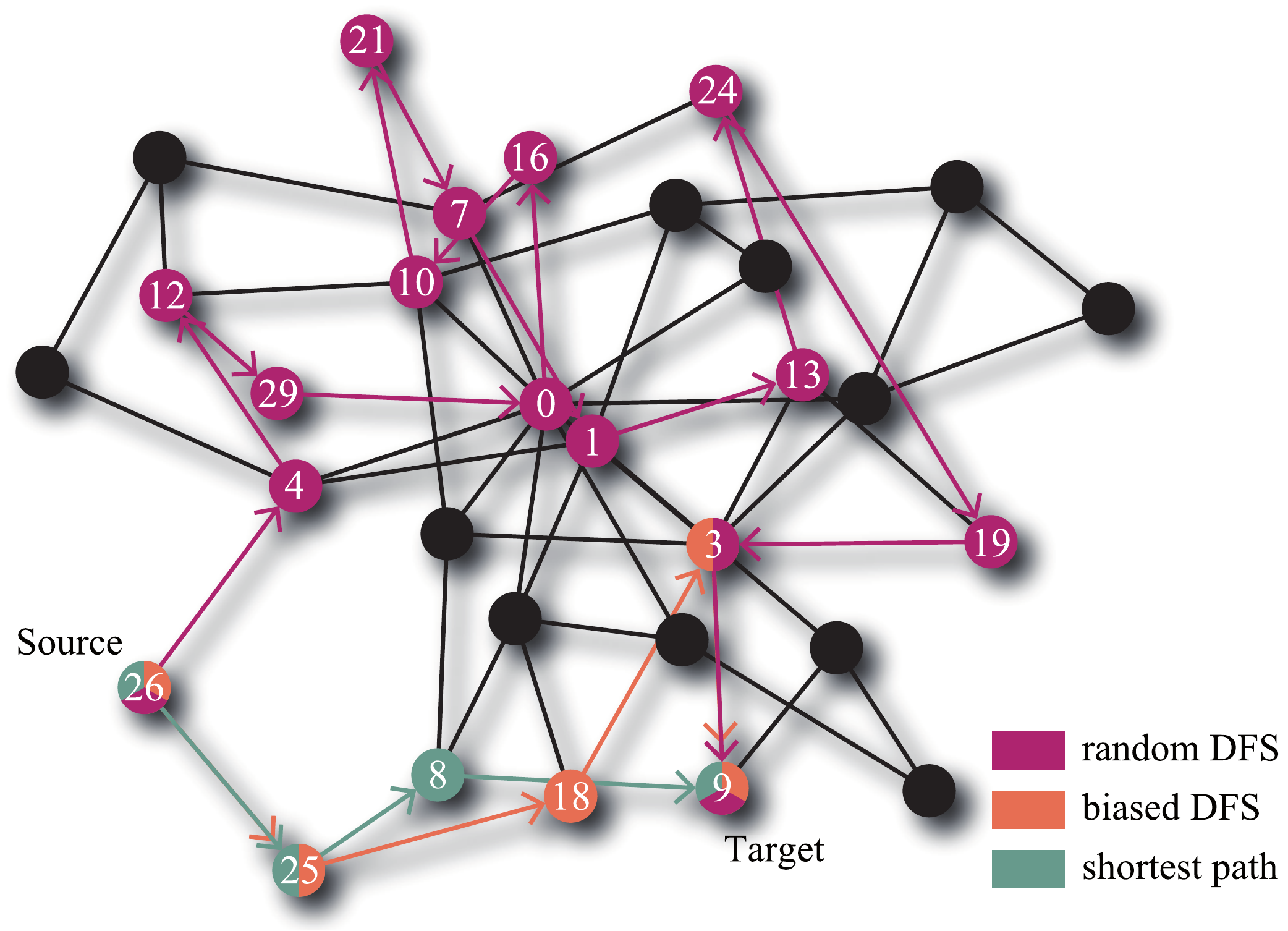}
\caption{Illustration of two geometric routing processes:
biased DFS, random DFS, and the shortest path.
Here the source vertex is $26$, and the agents aim to find
$9$. In bDFS, the packet selects the neighbor $18$ of the vertex $25$,
because it is geometrically closer
to the target. This causes a suboptimal routing of
$4$ steps,
instead of $3$ steps.
However, bDFS is more efficient than a random DFS realization of
$14$ steps
using no
geometric information.
}
\label{BA_DFS_routing_ex}
\end{center}
\end{figure}

Another way of assigning information to vertices than the indices of ASU and ASD, is to embed the vertices in space. In principle the methods are similar---the geographic information becomes a point in $\mathbb{R}^2$ rather than an integer index. If the agent knows the coordinates of the target it can steer toward it in Euclidean space just like in graph space.
In a recent paper, Bogu{\~n}{\'a} {\em et al}.
use geographic embedding of graph models as a basis for a routing strategy~\cite{Boguna2008,Boguna2008a}.
In real navigation, the knowledge about the direction
to the target is available even if  global information about
the structure of connection is unknown. If a graph's layout
in Euclidean space reflects the graph distances (so that a short Euclidean distance means a short graph distance) then the geographic information could be more valuable than the indices of the ASU and ASD schemes---one could just go toward the target as straight as the layout permits.

The geometric routing strategy we use in this work, for graphs embedded in
Euclidean space, makes a DFS of the target where every step down the search tree goes to the vertex closest  to the target. If there is no unvisited vertex available at a deeper level, the agent backtracks to the deepest level above with an available vertex to step down to.
Note that according to this rule, if a target vertex is
in the neighbor of a current vertex, the target vertex is selected
in the next step because the Euclidean distance to the target, zero, will be the lowest.
We call this navigation strategy
biased DFS (bDFS), and investigate its performance compared to
an unbiased random DFS. The routing strategies on a small graph
are illustrated in Fig.~\ref{BA_DFS_routing_ex}, where
the success of bDFS in Fig.~\ref{BA_DFS_routing_ex} is
more than just coincidences. The graph layout was done by the ``Neato'' layout of the Graphviz software (which is essentially the Kamada--Kawai algorithm~\cite{KK}). One of the goals of the  Kamada--Kawai layout is that vertices close in graph distance should be geometrically close too. Logically, moving closer in geometry should correspond to moving closer in graph space as well, which in turn helps the DFS routing protocol.
Again, we emphasize that the graph layout process itself can be time-consuming especially for large graphs, it is our main interest that the situation when the graph is used by individual agents with the pieces of information provided.

\subsection{Simulation results on model networks}

\begin{figure}
\begin{center}
\includegraphics[width=\textwidth]{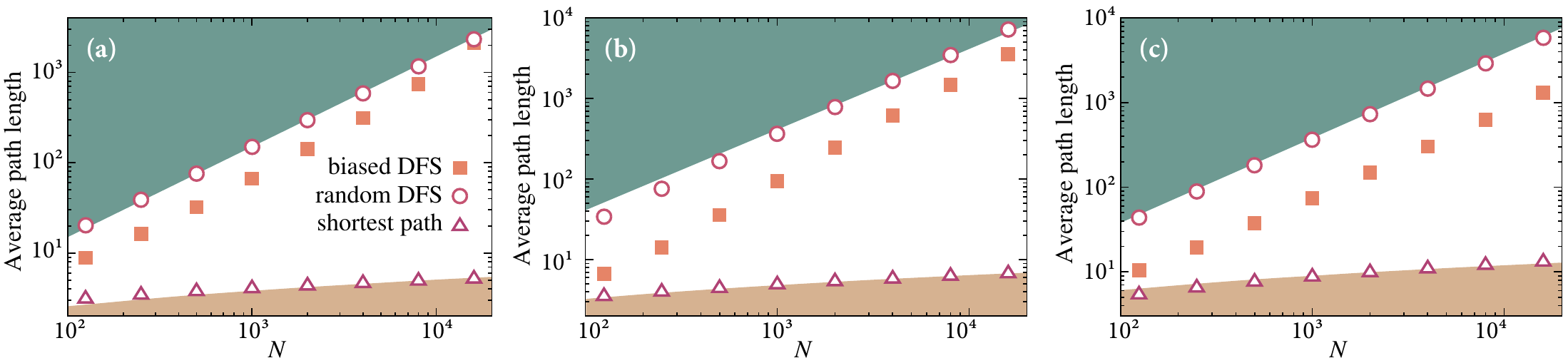}
\caption{Average navigation pathlengths (defined as number of hopping) for
geometric routing: bDFS vs random DFS. The upper and lower fields represent a linear and logarithmic scaling, respectively.
(a) BA model~\cite{Barabasi1999} with $m = 2$,
(b) HK model~\cite{Holme2002b}
with $m = 2$, (c) WS model~\cite{Watts1998} with $k = 4$ and $p = 0.1$.
All the error bars from standard error are smaller than the symbols.
}
\label{DFS_real}
\end{center}
\end{figure}

In Fig.~\ref{DFS_real}, we investigate the performance of bDFS relative to random DFS and the average distances (lengths of shortest paths). In this case we replace the ER model graphs by the Watts--Strogatz (WS) small-world network model~\cite{Watts1998}, to incorporate some geometric aspects. In this model, one starts by constructing a circulant graph (a circular graph where a vertex is connected to its $k$ nearest neighbors, symmetrically, on the circle). With a probability $p$, one edge is detached in one end and connected to a random other vertex.
We use up to $10^3$ graph realizations and up to $10^4$  $s\to t$ pairs.
Obviously, the bDFS strategy is substantially better, in all cases, than the random DFS. For the BA and HK graphs, however, the difference is decreasing with size.
This could perhaps be attributed to the layout program does not converge.
In this work we will not go into details of how the Neato or Kamada--Kawai algorithms work statistically, more than this observation---the navigational information they embed does not seem to scale with the system size.
For the WS graph, however, this difference is smaller, because the circulant graph, and thus the WS graph has a natural, spatially extended, layout (and is in this sense more ``geometrical'' than the BA and HK model graphs). Also, the clustering structure in HK graphs help the routing for better performance, as we can see from Figs.~\ref{DFS_real}(a) and (b). Therefore, the amount of embedded information exploitable is $\textrm{BA} < \textrm{HK} < \textrm{WS}$, clearly shown in Fig.~\ref{DFS_real}.

The graphviz layout, however, is not perfect for navigation. For instance, it is shown that the circular layout for WS small-world network model gives the searching time scaling as $\sqrt{N}$~\cite{Moura2003} if the greedy algorithm is used. We observe that the graphviz does not produce the circular layout which leads to the suboptimal (linear size dependence) performance shown in Fig.~\ref{DFS_real}(c).
Comparing Figs.~\ref{ASD_ASU_real} and \ref{DFS_real}, we see that the geographic routing is much less efficient than e.g.\ the ASD scheme, even though it encodes more information (two real numbers compared to one integer in $[1,N]$). Future studies are needed to see how far one can push the efficiency of geographic routing.

\section{Conclusions}
\label{conclusion}
In summary, we have investigated how agents can find their ways between vertices of a graph where incomplete information is added to the graph to facilitate navigation. We present two navigation schemes, ASU and ASD, where the indices of the vertices are chosen so that the agent can navigate half their way to the target (to the root of an embedded search tree). We show that by
increasing the information in indexing the vertices of a graph, so that each vertex can have two indices ASU and ASD can be combined to a much more efficient algorithm. We proceed to investigate the added value for navigation from graph layout programs. Such an algorithm, Neato from Graphviz, to add information to a DFS search, but not so much that such geographic navigation comes close to ASU or ASD. Our studies open questions for the future of how the  average shortest path (for an optimal algorithm for embedding and decoding the information) scales with the amount of distributed information. This is especially interesting for geographically embedded graphs---like sensor networks or other networks of wireless devices, perhaps even Internet---where this type of routing protocols potentially have applications in the future. In addition, more quantified measures of useful information for such routing processes such as entropy~\cite{Bianconi2009} would be worth investigating.

\section*{Acknowledgements}
This research was supported by the Swedish Research Foundation and the WCU program through NRF Korea funded by MEST R31--2008--000--10029--0 (PH).






\begin{thebibliography}{00}


\bibitem{Adamic2001} L.A. Adamic, R.M. Lukose, A.R. Puniyani,
B.A. Huberman, Phys. Rev. E 64 (2001) 046135.

\bibitem{Kleinberg2000} J.M. Kleinberg, Nature 406 (2000) 845.

\bibitem{Boguna2008} M. Bogu{\~n}{\'a}, D. Krioukov,
K.C. Claffy, Nat. Phys. 5 (2008) 74.

\bibitem{Boguna2008a} M. Bogu{\~n}{\'a}, D. Krioukov, Phys. Rev. Lett. 102 (2009)
058701.

\bibitem{MartinRosvall} K. Sneppen, A. Trusina, M. Rosvall,
Europhys. Lett. 69 (2005) 853;
M. Rosvall, P. Minnhagen,
K. Sneppen, Phys. Rev. E 71 (2005) 066111;
M. Rosvall, A. Gr{\"o}nlund, P. Minnhagen, K. Sneppen,
Phys. Rev. E 72 (2005) 046117.

\bibitem{Holme2007a} P. Holme,
Dynamics on and of networks, N. Ganguly, A. Mukherjee,
A. Deutsch, eds., Birkauser, 2009, pp. 189--198.

\bibitem{TraceRoute1} A. Clauset, C. Moore, Phys. Rev. Lett.
94 (2005) 018701.

\bibitem{TraceRoute2} L. Dall'Asta, I. Alvarez-Hamelin,
A. Barrat, A. V{\'a}zquez, A. Vespignani,
Phys. Rev. 71 (2005) 036135.

\bibitem{HYoun2007} H. Youn, M.T. Gastner, H. Jeong,
Phys. Rev. Lett. 101 (2008) 128701.

\bibitem{MANET} O.K. Tonguz, Ad Hoc Wireless Network:
A Communication-Theoretic Perspective,
Wiley, 2006.

\bibitem{WSN} I.F. Akyildiz, W. Su, Y. Sankarasubramaniam,
E. Cayirci, Computer Networks 38 (2002) 393.

\bibitem{JDNoh2004} J.D. Noh, H. Rieger, Phys. Rev. Lett.
92 (2004) 118701.

\bibitem{Newman2001} M.E.J. Newman, S.H. Strogatz, D.J. Watts,
Phys. Rev. E 64 (2001) 026118.

\bibitem{Kalisky2006} T. Kalisky, R. Cohen, O. Mokryn, D. Dolev,
Y. Shavitt, S. Havlin, Phys. Rev. E 74 (2006) 066108.

\bibitem{Barabasi1999} A.-L. Barab{\'a}si, R. Albert, Science 286 (1999)
509.

\bibitem{Holme2002b} P. Holme, B.J. Kim, Phys. Rev. E 65 (2002)
026107.

\bibitem{Erdos1959} P. Erd\H{o}s, A. R{\'e}nyi, Publ. Math.
6 (1959) 290.

\bibitem{Cohen2003} R. Cohen, S. Havlin, Phys. Rev. Lett. 90 (2003)
058701.

\bibitem{Bianconi2009} G. Bianconi, Phys. Rev. E 79 (2009)
036114.

\bibitem{KIGoh2003} K.-I. Goh, E. Oh, B. Kahng, D. Kim,
Phys. Rev. E 67 (2003) 017101.

\bibitem{KK} T. Kamada, S. Kawai,
Information Processing Letters 31 (1989) 7.

\bibitem{Watts1998} D.J. Watts, S.H. Strogatz, Nature 393 (1998)
440.

\bibitem{Moura2003} A.P.S. de Moura, A.E. Motter, C. Grebogi,
Phys. Rev. E 68 (2003) 036106.

\end{thebibliography}



\end{document}